\documentclass[twocolumn,showpacs,preprintnumbers,amsmath,amssymb,nofootinbib]{revtex4}
\usepackage{mathbbold}
\usepackage{amsmath}
\usepackage{amssymb}
\usepackage{graphicx}
\usepackage{dcolumn}
\usepackage{bm}

\begin{document}

\title{Exploring the structural regularities in networks}
\author{Hua-Wei Shen}
\email{shenhuawei@ict.ac.cn}
\author{Xue-Qi Cheng}
\email{cxq@ict.ac.cn}
\author{Jia-Feng Guo}
\affiliation{Institute of Computing Technology, Chinese Academy of
Sciences, Beijing 100190, China}

\date{\today}

\begin{abstract}
In this paper, we consider the problem of exploring structural
regularities of networks by dividing the nodes of a network into
groups such that the members of each group have similar patterns of
connections to other groups. Specifically, we propose a general
statistical model to describe network structure. In this model,
group is viewed as hidden or unobserved quantity and it is learned
by fitting the observed network data using the
expectation-maximization algorithm. Compared with existing models,
the most prominent strength of our model is the high flexibility.
This strength enables it to possess the advantages of existing
models and overcomes their shortcomings in a unified way. As a
result, not only broad types of structure can be detected without
prior knowledge of what type of intrinsic regularities exist in the
network, but also the type of identified structure can be directly
learned from data. Moreover, by differentiating outgoing edges from
incoming edges, our model can detect several types of structural
regularities beyond competing models. Tests on a number of real
world and artificial networks demonstrate that our model outperforms
the state-of-the-art model at shedding light on the structural
features of networks, including the overlapping community structure,
multipartite structure and several other types of structure which
are beyond the capability of existing models.
\end{abstract}


\maketitle

\section{Introduction}
Network provides a powerful tool for representing the structure of
complex systems. These networks include social
networks~\cite{Girvan02,Newman03a}, information
networks~\cite{Flake02, Cheng09}, and biological
networks~\cite{Girvan02,Guimera05}. Much of recent research on
networks actually aims to understand the structural regularities and
further to reveal the relationship between such structural
regularities and the function of networks. For example, as a
widely-studied structural characteristic of network, community
structure is of high interest because communities often correspond
to functional units such as pathways for metabolic networks and
collections of pages on a similar topic on the Web.

Community structure is a kind of assortative structure, in which
nodes are divided into groups such that the members within each
group are mostly connected with each other. Contrary to community
structure, multipartite structure is another important kind of
structural regularities observed in real world networks.
Multipartite structure means that nodes of network can be divided
into groups such that most of edges are across groups. Beside these
salient structural characteristics, other types of structure are
also observed in real world networks, such as hierarchical structure
and core-periphery structure.

However, existing methods mostly presume that certain type of
structure exists in the target network and then devote to detect
such structure. This raises concerns to the reliability of the
resulted structure. On one hand, the assumed structure may not match
the intrinsic structure of the target network and thus these methods
are not applicable to these situations. On the other hand, several
real world networks contain multiple types of structure
simultaneously. Most existing methods are designed for certain type
of structure and thus cannot detect the broad types of structure. In
addition, several unknown types of structure may also exist in
networks and a desired method should be able to detect such
structure as well. Thus, it is the time to explore multiple types of
structural regularities in networks.

In the last decade, the identification of community structure has
attracted much attention in various scientific fields. Many methods
have been proposed and applied successfully to some specific complex
networks~\cite{Newman04,Clauset04,Radicchi04,Palla05,Duch05,Newman06a,
Newman06b,Shen09a,Shen09b,Cheng10,Shen10}. For review, the reader
can refer to Ref.~\cite{Fortunato10}. These methods are from
different perspectives, such as the centrality measures, modularity,
link density, percolation theory, network compression, and spectral
analysis. Recently, several generative models for network data are
proposed to detect community structure~\cite{Ramasco08,Vazquez08}.
These models view network structure as observed quantities and take
communities as hidden groups of nodes. The communities are then
identified by fitting the model to the observed network structure.
For example, Ren \textit{et al.}~\cite{Ren09} proposed a
probabilistic model to uncover the overlapping community structure.
This model assumes that the two end nodes of each edge are from the
same community and this assumption is satisfied by the fuzzy
membership of nodes. Zhang \textit{et al.}~\cite{Zhang07} applied
the Latent Dirichlet Allocation (LDA, a well-known generative model)
to social network analysis and gave a method to detect community
structure. The common drawback of these two models is that they can
only uncover the community structure and fail to reveal other types
of structural regularities, e.g., multipartite structure.

To characterize the hierarchical organization of networks,
Clauset~\textit{et al.} proposed the hierarchical random graph
model, which is capable of expressing both assortative and
disassortative structure~\cite{Clauset08}. To explore more broad
types of structure, Newman \textit{et al.} proposed a mixture model
for exploratory analysis of network structure~\cite{Newman07}. In
this model, the nodes with similar connection preference rather than
the highly connected nodes are classified into the same group. In
such a general way, this model can reveal several other kinds of
structural regularities beyond community structure. However, this
model fails to tell us which kind of structural regularities has
been identified. More importantly, this model may produce a result
which is a mixture of several types of structure, and thus the
identified structure may not provide clear information about the
structural regularities. The shortcoming of this model is attributed
to that it only models the relationship between groups and nodes
rather than the relationship among groups. Stochastic blockmodel
provides an appropriate alternative to the mixture model for
exploring broad range of structural regularities. Karrer \textit{et
al.} utilized a degree-corrected stochastic
blockmodel~\cite{Karrer11} to investigate community structure of
network. Airoldi \textit{et al.} gave a mixed membership stochastic
blockmodel~\cite{Airoldi08} to model network data. These work has
demonstrated that stochastic blockmodel is a good choice for
exploring regularities of network. However, the effectiveness of
these models are limited by their inflexible model assumptions,
e.g., the hard partition assumption or neglecting the directionality
of edges.

In this paper, we focus on exploring the intrinsic structural
regularities in network by dividing network nodes into groups such
that the members of each group have similar patterns of connections
to other groups. We propose a general stochastic blockmodel
(referred to as GSB model in this paper) to model the network
structure. In this model, node groups are represented by unobserved
or hidden quantities and the relationship among groups are
explicitly modeled by a block matrix as the traditional blockmodels.
Then, using the expectation-maximization algorithm, we fit the model
to specific network data and thus detect intrinsic structural
regularities of the network without prior knowledge of what type of
regularity exists in the network. Compared with existing models, the
most prominent strength of our model is the high flexibility. This
strength enables it to possess the advantages of existing models and
overcomes their shortcomings in a unified way. As a result, not only
broad types of structure can be detected, but also the type of
identified structure can be indicated by the block matrix. In
addition, our model can tell us the centrality of the node in each
group and the mixed membership of nodes as well.

Tests on a number of artificial and real world networks demonstrate
that our model outperforms the state-of-the-art models at shedding
light on the structural features of networks, including the
overlapping community structure, multipartite structure and several
other types of structure which is beyond the capability of existing
models.

\section{The Model}
\label{sec2}

Generally, a network with $n$ nodes can be represented
mathematically by an adjacency matrix $A$ with elements $A_{ij}=1$
if there is an edge from node $i$ to node $j$ and $0$ otherwise. For
weighted networks, $A_{ij}$ is generalized to represent the weight
of the edge from $i$ to $j$.

To investigate the structural regularities in network, we suppose
that the $n$ nodes of the network fall into $c$ groups whose
memberships are unknown, i.e., we cannot observe or measure them
directly. In this paper, we propose a statistical model to infer the
group membership from the observed network structure.

The model we used is a kind of stochastic blockmodel. Blockmodel is
a generative model and has a long tradition of study in the social
science and computer science. For a standard blockmodel, a $c\times
c$ matrix $\omega$ is generally adopted such that the matrix element
$\omega_{rs}$ denotes the probability that a randomly selected edge
connects group $r$ to group $s$, i.e., the tail node of the edge is
from group $r$ and the head node is from $s$. The advantage of
blockmodel lies in that the matrix $\omega$ explicitly characterizes
various types of connecting patterns among groups.

In the standard blockmodel, the nodes in the same group are
identical, i.e., each node in a group has equal probability to be
the end node of an edge adjacent to the group. This constraint is
relaxed in our model. Specifically, for an edge with its tail node
being from group $r$ and its head node being from group $s$, we use
$\theta_{ri}$ to denote the probability that the tail node is $i$
and $\phi_{sj}$ to denote the probability that the head node is $j$
respectively. In addition, we use $\overrightarrow{g}_{ij}$ and
$\overleftarrow{g}_{ij}$ to denote respectively the group membership
of the tail node and head node of the edge $e_{ij}$.

Up to now, we have given all the quantities in our model. They can
be classified into three classes: observed quantities $\{A_{ij}\}$,
hidden quantities $\{\overrightarrow{g}_{ij},
\overleftarrow{g}_{ij}\}$, and model parameters $\{\omega_{rs},
\theta_{ri}, \phi_{sj}\}$. To simplify the notations, we henceforth
denote by $A$ the entire set $\{A_{ij}\}$ and similarly
$\overrightarrow{g}$, $\overleftarrow{g}$, $\omega$, $\theta$,
$\phi$ for $\{\overrightarrow{g}_{ij}\}$,
$\{\overleftarrow{g}_{ij}\}$, $\{\omega_{rs}\}$, $\{\theta_{ri}\}$
and $\{\phi_{sj}\}$.

With our model, an edge $e_{ij}$ is generated in the following
process:
\begin{enumerate}
\item
Select two groups $\overrightarrow{g_{ij}}=r$ and
$\overleftarrow{g_{ij}}=s$ respectively for the tail node and head
node of the edge with probability $\omega_{rs}$;
\item
Draw the tail node $i$ from the group $r$ with probability
$\theta_{ri}$;
\item
Draw the head node $j$ from the group $s$ with probability
$\phi_{sj}$.
\end{enumerate}
Summing over the latent quantities $r$ and $s$, the probability that
we observe an edge $e_{ij}$ can be written as
\begin{equation}
\text{Pr}(e_{ij}|\omega,\theta,\phi)=\sum_{rs}\omega_{rs}\theta_{ri}\phi_{sj}.\label{eq1}
\end{equation}

Then, the likelihood of the observed network with respect to our
model is
\begin{equation}
\text{Pr}(A|\omega, \theta,
\phi)=\prod_{ij}\left(\sum_{rs}\omega_{rs}\theta_{ri}\phi_{sj}\right)^{A_{ij}}.\label{eq2}
\end{equation}
Note that the self-loop edges are allowed and the weight $A_{ij}$ is
taken as the number of multi-edges connecting node $i$ to node $j$
as done in many existing models including, for instance, the widely
studied configuration model~\cite{Newman01}.

Intuitively, the parameter $\theta_{ri}$ characterizes the
centrality of node $i$ in the group $r$ from the perspective of
outgoing edges while $\phi_{sj}$ describes the centrality of node
$j$ in the group $s$ from the perspective of incoming edges.
Differently from traditional blockmodels, by differentiating these
two kinds of centrality, our model can provide more flexibility to
explore broad types of intrinsic structural regularities in network.
Note that the parameters $\omega_{rs}$, $\theta_{ri}$, $\phi_{sj}$
satisfy the normalization conditions
\begin{eqnarray}
\sum_{r=1}^{c}\sum_{s=1}^{c}\omega_{rs}=1,\hspace{2mm}\sum_{i=1}^{n}\theta_{ri}=1,
\hspace{2mm}\sum_{j=1}^{n}\phi_{sj}=1. \label{eq3}
\end{eqnarray}

Now our task is to estimate the model parameters and to infer the
unobserved quantities by fitting the model to the observed network
data. The standard framework for such a task is likelihood
maximization. Generally, one works not with the likelihood
[\,Eq.~(\ref{eq2})\,] itself but with its logarithm (log-likelihood)
\begin{eqnarray}
\mathcal{L}&=& \text{ln}\text{Pr}(A|\omega, \theta,
\phi)\nonumber \\
&=&
\sum_{ij}{A_{ij}\text{ln}\left(\sum_{rs}\omega_{r,s}\theta_{ri}\phi_{sj}\right)}
.\label{eq4}
\end{eqnarray}
The maximums of the likelihood and its logarithm are in the same
place since the logarithm is a monotonically increasing function.

Directly maximizing the log-likelihood is difficult because of the
inner sum over the unobserved quantities $\overrightarrow{g}_{ij}=r$
and $\overleftarrow{g}_{ij}=s$. Using Jensen's inequality, the
maximization of the log-likelihood can be transformed to the
maximization of the expected log-likelihood
\begin{eqnarray}
\overline{\mathcal{L}}&=&\sum_{\overrightarrow{g},\overleftarrow{g}}\text{Pr}(\overrightarrow{g},\overleftarrow{g}|A,\omega,\theta,\phi)\text{ln}\text{Pr}(A|\overrightarrow{g},\overleftarrow{g}, \omega, \theta, \phi)\nonumber\\
&=&\sum_{ijrs}\text{Pr}(\overrightarrow{g}_{ij}=r,\overleftarrow{g}_{ij}=s|e_{ij},\omega,\theta, \phi)\nonumber\\
&&\hspace{5.5mm}\bigg[A_{ij}\big(\text{ln}\omega_{rs}+\text{ln}\theta_{ri}+\text{ln}\phi_{sj}\big)\bigg]
\nonumber \\
&=&\sum_{ijrs}{q_{ijrs}A_{ij}\left(\text{ln}\omega_{rs}+\text{ln}\theta_{ri}+\text{ln}\phi_{sj}\right)},\label{eq5}
\end{eqnarray}
where to simplify the notation we have defined
$q_{ijrs}=\text{Pr}(\overrightarrow{g}_{ij}=r,\overleftarrow{g}_{ij}=s|e_{ij},\omega,\theta,\phi)$,
which denotes the probability that one observes an edge $e_{ij}$
with its tail node $i$ from group $r$ and its head node $j$ from
group $s$ given the observed network and the model parameters.

With the expected log-likelihood, we can give the best estimate of
the value $\overline{\mathcal{L}}$ and the position of its maximum
represents the best estimate of the most likely values of the model
parameters. Specifically, if the value of $q_{ijrs}$ is known, we
can find the values of the model parameters $\omega$, $\theta$,
$\phi$ where $\overline{\mathcal{L}}$ reaches its maximum. However,
the calculation of $q_{ijrs}$ requires the values of these model
parameters. To address such a problem, an expectation-maximization
(EM) algorithm is adopted.

Under the framework of EM algorithm, we first calculate the value of
$q_{ijrs}$ by
\begin{eqnarray}
q_{ijrs}&=&\frac{\text{Pr}(\overrightarrow{g}_{ij}=r,\overleftarrow{g}_{ij}=s,e_{ij}|\omega,\theta,\phi)}{\text{Pr}(e_{ij}|\omega,\theta,\phi)}\nonumber\\
&=&\frac{\omega_{rs}\theta_{ri}\phi_{sj}}{\sum_{rs}{\omega_{rs}\theta_{ri}\phi_{sj}}}.\label{eq6}
\end{eqnarray}
Once we have the values of the $q_{ijrs}$, we can use them to
evaluate the expected log-likelihood and hence to find the values of
$\omega$, $\theta$, $\phi$ that maximize it.

Introducing the Lagrange multipliers $\rho$, $\gamma_{r}$ and
$\lambda_{s}$ to incorporate the normalization conditions in
Eq.~(\ref{eq3}), the expected log-likelihood expression to be
maximized becomes
\begin{eqnarray}
\tilde{\mathcal{L}}&=&\overline{\mathcal{L}}+\rho\bigg(1-\sum_{rs}{\omega_{rs}}\bigg)+\sum_{r}{\gamma_{r}\bigg(1-\sum_{i}{\theta_{ri}}\bigg)}
+\nonumber \nonumber  \\ &&
\sum_{s}{\lambda_{s}\bigg(1-\sum_{j}{\phi_{sj}}\bigg)}.\label{eq7}
\end{eqnarray}

By letting the derivative of $\tilde{\mathcal{L}}$ to be $0$, the
maximum of the expected log-likelihood occurs at the places where

\begin{equation}
\label{eq8} \left\{
\begin{aligned}
\omega_{rs}&=\frac{ \sum_{ij}{A_{ij}q_{ijrs}} } { \sum_{ijrs}{A_{ij}q_{ijrs}}  }, \\
\theta_{ri} &= \frac{\sum_{js}{A_{ij}q_{ijrs}}}{\sum_{ijs}{A_{ij}q_{ijrs}}}, \\
\phi_{sj} &=
\frac{\sum_{ir}{A_{ij}q_{ijrs}}}{\sum_{ijr}{A_{ij}q_{ijrs}}}.
\end{aligned} \right.
\end{equation}

Eqs.~(\ref{eq6}) and~(\ref{eq8}) constitute our
expectation-maximization algorithm. In the expectation step, the
expected value of log-likelihood is calculated through evaluating
the values of $q_{ijrs}$ with Eq.~(\ref{eq6}). In the maximization
step, the expected value of log-likelihood is maximized when the
values of model parameters $\omega$, $\theta$, $\phi$ are evaluated
with Eq.~(\ref{eq8}). Implementation of the algorithm consists
merely of iterating Eqs.~(\ref{eq6}) and~(\ref{eq8}) until
convergence.

When the algorithm converges, we obtain a set of values for hidden
quantity $q_{ijrs}$ and model parameters $\omega$, $\theta$, $\phi$.
This set of values is self-consistent with respect to
Eqs.~(\ref{eq6}) and~(\ref{eq8}). However, it is not always the
place where the log-likelihood reaches its maximum. In other words,
the expectation-maximization algorithm may converge to local maxima
of the log-likelihood. With different starting values, the algorithm
will give rise to different solutions. To obtain a satisfactory
solution, it is necessary to perform several runs with different
initial conditions and take the solution giving the highest
log-likelihood over all the runs performed.

By fitting the model to the observed network structure with the
expectation-maximization algorithm, the estimated model parameters
provide us vital information for structural regularities of the
network. Specifically, $\theta$ and $\phi$ describe the centrality
of a node in groups containing it respectively from the perspective
of outgoing edges and incoming edges. The parameter $\omega$
characterizes the connecting patterns among different groups, i.e.,
the type of structural regularities.

More importantly, according to the model parameters, we can define
two kinds of group memberships $\alpha_{ir}$ and $\beta_{js}$
respectively from the perspective of outgoing edges and incoming
edges. Specifically, $\alpha_{ir}$ is the probability that node $i$
is from group $r$ when it acts as the tail node of edges while
$\beta_{js}$ is the probability that node $j$ is from group $s$ when
it acts as the head node of edges. For $\alpha_{ir}$, it can be
calculated by
\begin{equation}
\alpha_{ir}=\frac{\sum_{s}{\omega_{rs}\theta_{ri}}}{\sum_{rs}{\omega_{rs}\theta_{ri}}}.\label{eq9}
\end{equation}
Actually, $\alpha_{ir}$ provides a soft or fuzzy membership, i.e.,
node $i$ can belong to more than one groups simultaneously. When the
identified structural regularity corresponds to community structure,
we actually obtain the overlapping community structure which has
attracted much research attention ever since it is proposed. If one
wants to get a hard partition, we can simply assign each node $i$ to
the group $r$ satisfying $r=\text{arg }\text{max}_{s}\{\alpha_{is},
s=1,2,\cdots,c\}$. These statements for $\alpha_{ri}$ also apply to
$\beta_{ir}$ defined as
\begin{equation}
\beta_{js}=\frac{\sum_{r}{\omega_{rs}\phi_{sj}}}{\sum_{rs}{\omega_{rs}\phi_{sj}}}.\label{eq10}
\end{equation}

Finally, the model described above so far is based on directed
networks. Actually, the model can be easily generalized to
undirected networks by letting the parameter $\theta$ be identical
to $\phi$. The derivation follows the case of directed networks and
the results are the same to Eqs.~(\ref{eq6}) and~(\ref{eq8}).

Now we discuss the computational cost of the
expectation-maximization algorithm for the fitting of our model. For
each iteration in this algorithm, the cost consists in two parts.
The first part is from the calculation of $q_{ijrs}$ using
Eq.~(\ref{eq6}), whose time-complexity is $O(m\times c^2)$. Here $m$
is the edges in the network and $c$ is the number of groups. The
second part is from the estimation of the model parameters using
Eq.~(\ref{eq8}), whose time-complexity is also $O(m\times c^2)$. We
use $T$ to denote the number of iterations before the iteration
process converges. Then, the total cost of the
expectation-maximization algorithm for our model is $O(T\times m
\times c^2)$. It is difficult to give a theoretical estimation to
the number $T$ of iterations. Generally speaking, $T$ is determined
by the network structure and the initial condition. A well-designed
initial condition may result in good convergence rate. The
computational cost limits our model to deal with networks with
moderate scale networks. We look forward to seeing more efficient
implementation for our model. Note that the method proposed
in~\cite{Ball11} provides a promising way to improve the
computational efficiency and decrease the memory space required.

\section{Comparison with other models}
\label{sec3}

\begin{figure}
\centering
\includegraphics[width=0.44\textwidth]{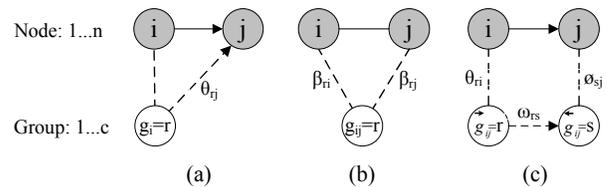} \caption{Generative models for network data:
(a) Newman's mixture model~\cite{Newman07}, (b) Probabilistic model
proposed in~\cite{Ren09}, and (c) our model. Filled circles
represent observed quantities and unfilled ones correspond to hidden
quantities. The solid line (with arrow) between node $i$ and $j$
indicates the existence of one (directed) edge connecting them. The
dashed-line connecting two circles indicates that the relation
between the corresponding quantities is unobserved and requires
being learned from the observed network data. Arrows represent the
directions of relation.} \label{fig1}\vspace{-10pt}
\end{figure}

In this section, we illustrate the difference and connections
between our model with several existing models. Fig.~\ref{fig1}
gives the schematic for our model and two existing generative
models, namely Newman's mixture model and Ren's model.

\begin{figure}
\centering
\includegraphics[width=0.44\textwidth]{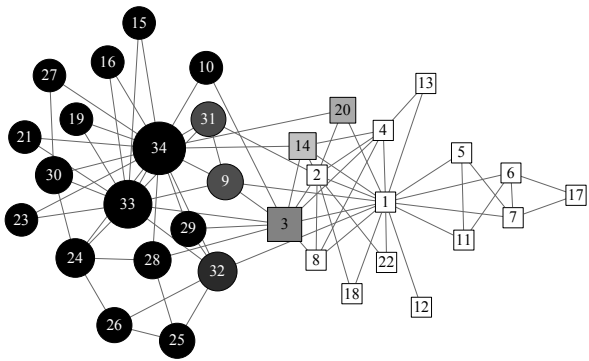} \caption{The network of the
karate club studied by Zachary~\cite{Zachary77}. The real social
fission of this network is represented by two different shapes,
circle and square. The shades of nodes indicate the mixed membership
obtained by fitting our model to this network. The sizes of the
nodes indicate the centrality degree of nodes with respect to the
left group, i.e., $\theta_{ri}$ ranging from $0$ for the smallest
nodes to $0.22$ for the largest.} \label{fig2} \vspace{-10pt}
\end{figure}

For Newman's model, as shown in Fig.~\ref{fig1}(a), each group $r$
is characterized by the connecting preference $\theta_{rj}$ to node
$j$, no matter the node $j$ is contained by the group $r$ or not.
The nodes belonging to the same group have similar connecting
preference. As a result, both assortative and disassortative
structural regularities can be detected by this model. However, this
model has no parameter to explicitly characterize the type of the
identified structure. More importantly, this model may produce a
result which is a mixture of several types of structure and thus in
these cases the identified structure may provide confused
information about the structural regularities. For example, for the
network shown in Fig.~\ref{fig2}, nodes $12$, $15$, $16$, $19$,
$21$, $23$ are identified by this model as overlapped nodes shared
by the two groups, denoted by circles and squares, although these
nodes only have connections to one of the two groups.

For Ren's model, as shown in Fig.~\ref{fig1}(b), the two end nodes
of each edge is assumed to be from the same group. As a result, only
the assortative structure (community) can be detected using this
model. Note that, for this model, no edge is allowed to connect
different groups. The relationship between communities is instead
reflected by the overlapped nodes.

For our model, it essentially is a kind of stochastic blockmodel, in
which the relationships among different node groups are explicitly
modeled by the block matrix $w$. In this way, our model possesses
the advantages of both Newman's model and Ren's model and overcomes
the shortcoming of these two models.

On one hand, through learning the matrix $w$ according to observed
network data, various types of structural regularities can be
explored by our model. The type of the identified structure is
indicated by the matrix $w$. Specifically, when the matrix $\omega$
is an identity matrix, the identified structural regularity
corresponds to an obvious community structure. Meanwhile,
multipartite or anti-community structure is revealed when the
estimated model parameter $\omega$ is an anti-diagonal matrix with
all the anti-diagonal elements being $1$. For other types of
structure such as core-periphery structure and hierarchical
structure, the form of $\omega$ is the same to the block matrix
$\omega$ in traditional block models~\cite{Karrer11}.

On the other hand, using the matrix $w$, our model discards the
assumption of Ren's model that two end nodes of one edge are
required to be from the same community. In this sense, Ren's model
is a special case of our model. In addition, our model also provides
several other flexibility. By representing the centrality of nodes
in group from two different perspective respectively according to
the outgoing edges and incoming edges, our model can detect more
broad range of structural regularities which is out of the
capability of other models. This will be shown later in the
subsequent section. Moreover, our model can be further generalized
by not requiring the matrix $w$ be a square matrix.

Finally, we compare our model to two recently proposed stochastic
models for community detection~\cite{Karrer11,Ball11}. Firstly, both
our model and Karrer's model~\cite{Karrer11} are stochastic block
model where a block matrix is adopted to characterize the connecting
patterns among groups. The main difference between these two models
lies in that Karrer's model is designed to detect disjoint
structural regularities while our model is for fuzzy structural
regularities. This difference is reflected by the definition of the
model parameters $\theta$ and $\phi$ in our model and the definition
of the model parameter $\theta$ in Karrer's model. In addition, our
model differentiates the out-edges from in-edges of nodes while
Karrer's model does not. Secondly, similar to Ren's model, Ball's
model~\cite{Ball11} focused on the community structure while our
model can uncover multiple types of structural regularities.

\section{Experimental Results}

In this section, we demonstrate the effectiveness of our model at
exploring the structural regularities of networks by experiments on
several real world or artificial networks with various types of
intrinsic structural regularities. Then we discuss the model
selection issue, i.e., how to determine the optimal number of
groups.

\subsection{Detecting community structure}

The test network is the famous karate club network constructed by
Zachary. This network characterizes the acquaintance relationship
between $34$ members of a karate club in an American University. A
dispute arose between the club's administrator and its principal
karate teacher, and as a result the club eventually split into two
smaller clubs, centered around the administrator and the teacher
respectively. The network and its fission are depicted in
figure~\ref{fig2}. The administrator and the teacher are represented
by nodes $1$ and $33$ respectively.

By setting the group number $c=2$, we fit our model to the karate
club network data. The resulted matrix $\omega$ is a $2\times2$
identity matrix, indicating that the obtained structure is community
structure. Fig.~\ref{fig2} shows the two groups found by our model
with the expectation-maximization method. As shown in
Fig.~\ref{fig2}, the shades of the nodes in the figure represent the
values of $\alpha_{i1}$\footnote{Since this network is an undirected
network, the two kinds of belonging coefficient are identical, i.e.,
$\alpha_{ir}=\beta_{ir}$.}, where group $1$ is the left group. As we
can see, our model assigns most of the nodes strongly to one group
or the other. Actually, all but $6$ nodes are assigned $100\%$ to
one of the groups (black and white nodes in the figure). If we
simply divide the nodes into two disjoint groups by assigning each
node $i$ to the group $r$ according to the belong coefficients
$\alpha_{ir}$, the resulting groups perfectly correspond to the real
split of the club.

In addition, Table~\ref{tab1} gives the belonging coefficient of the
$6$ overlapped nodes which are shared by the two groups. These
overlapped nodes are nodes $3$, $9$, $14$, $20$, $31$, $32$. Note
that these overlapped nodes are often misclassified by traditional
partition-based community detection methods. For comparison, we also
gives the mixed membership of these six nodes according to Newman's
mixture model and Ren's model. As we can see, our model and Ren's
model produce the same results, which is attributed to the fact that
Ren's model is a special case of our model. However, Newman's model
behave very differently from the other two models. Actually, for
Newman's model, another $10$ nodes are also assigned to two groups,
e.g., nodes $12$, $15$. Such a result is counterintuitive to the
real structure of this network. As a conclusion, our model performs
better than Newman's model at detecting the overlaps between groups.
Ren's model can only detect community structure while our model can
detect other types of structural regularities as illustrated in the
following test.

\begin{table}
\centering \caption{\label{tab1} Mixed membership of overlapped
nodes.} \vspace{5pt}
\begin{tabular}{p{1.8cm}p{1.8cm}p{1.8cm}p{1.8cm}}
\hline \hline
Node ID & $\alpha_{i1}$ & $q_{i1}\footnotemark[1]$ & $\frac{u_{1i}}{u_{1i}+u_{2i}}\footnotemark[2]$\\
\hline
3  & 0.49 &  0.00 & 0.49\\
9  & 0.70 &  0.96 & 0.70\\
14 & 0.24 &  0.00 & 0.24\\
20 & 0.33 &  0.13 & 0.33\\
31 & 0.71 &  0.92 & 0.71\\
32 & 0.83 &  1.00 & 0.83\\
\hline\hline
\end{tabular}
\end{table}
\footnotetext[1]{$q_{ir}$ is defined in~\cite{Newman07} as the
probability that node $i$ belongs to group $r$.}
\footnotetext[2]{$\frac{u_{1i}}{u_{1i}+u_{2i}}$ is defined
in~\cite{Ren09} as the probability that node $i$ belongs to group
$r$.}

\subsection{Detecting multipartite structure}

\begin{figure}
\includegraphics[width=0.46\textwidth]{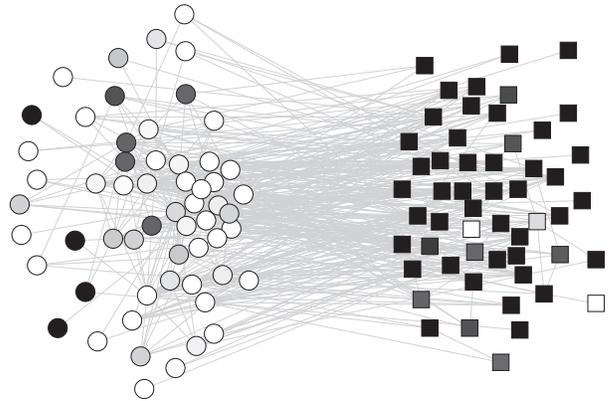} \caption{The adjacency network of English
words. Node groups corresponding to adjectives and nouns are
respectively denoted by circle and square. The shades of nodes
indicate their belonging coefficient obtained by fitting our model
to this network.} \label{fig3}\vspace{-10pt}
\end{figure}

Now we illustrate the detection of multipartite or anti-community
structure according to our model. The test network is the adjacency
network of English words taken from Ref.~\cite{Newman06b}. In this
network, the nodes represent $112$ commonly occurring adjectives and
nouns in the novel \textit{David Copperfield} by Charles Dickens,
with edges connecting any pair of words that appear adjacent to each
other at any place in the text. Generally, adjectives occur next to
nouns in English. Thus most edges in the network connect an
adjective to a noun and the network is approximately bipartite,
i.e., this network possesses anti-community structure. This can be
seen clearly in Fig.~\ref{fig3}, where the adjectives and nouns are
respectively represented by circles and squares.

Fitting our model to this network with $c=2$, the resulted $\omega$
is a transposed $2\times2$ identity matrix, indicating that the
identified structure is bipartite structure. The obtained two groups
and node memberships are shown by the shades of nodes as shown in
Fig.~\ref{fig3}. We can see that most nodes are assigned to only one
group, although there are several ambiguous cases corresponding to
the nodes with intermediate shades. If we assign each node to its
most preferred group, the resulted two disjoint groups well separate
the adjectives from the nouns. In fact, $100$ of all the $112$ nodes
are correctly classified. This accuracy is the same to the result
given by Newman's mixture model.

As a comparison, we also apply Ren's model to this network by
setting the group number being $2$. Only $60$ nodes of all the $112$
nodes are correctly classified, similar to the accuracy of random
assignment. The ineffectiveness of Ren's model at this network is
attributed to that Ren's model presumes the existence of community
structure in the network while the intrinsic structural regularity
is bipartite structure.

\subsection{Exploring other type of structural regularity}

In the previous tests, we have demonstrate that our model can be
used to detect both the assortative structure (i.e., community
structure) and the disassortative structure (i.e., multipartite
structure) without being told that which type of structural
regularities exists in the target networks. Now we will further show
that our model can also detect other type of structure which can not
be revealed by competing models.

\begin{figure}
  \begin{minipage}[t]{0.4\linewidth}
    \centering
    \includegraphics[width=0.9\textwidth]{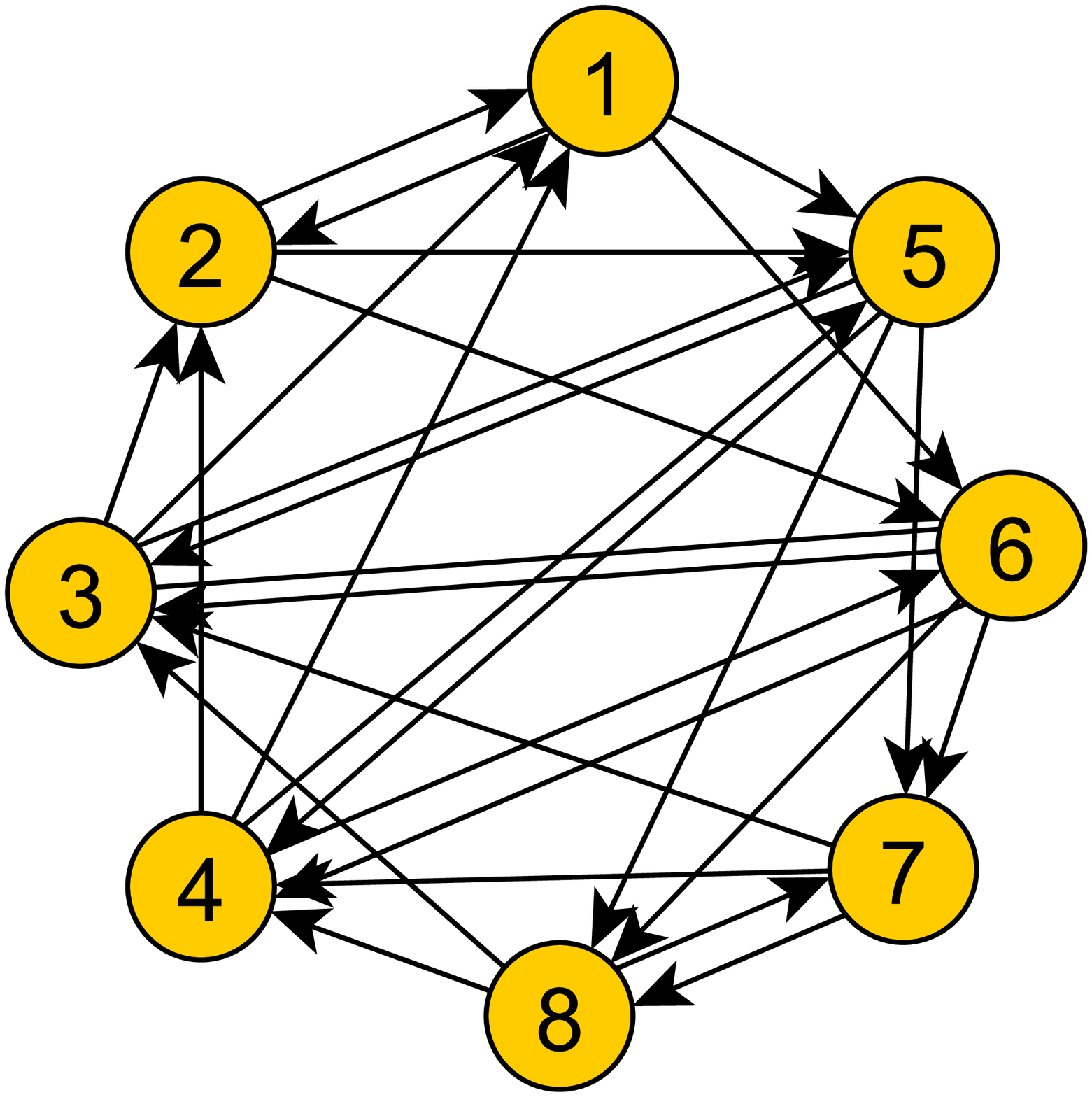}
\\ \hspace{5pt}\mbox{\rule{0pt}{8pt}{\rm (a)}}
  \end{minipage}%
  \begin{minipage}[t]{0.6\linewidth}
    \centering
    \includegraphics[width=0.7\textwidth]{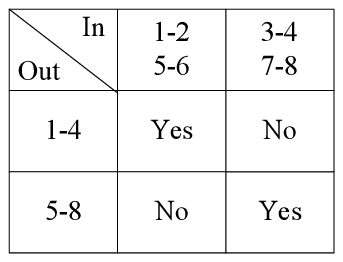}
\\ \hspace{10pt}\mbox{\rule{0pt}{8pt}{\rm (b)}}
  \end{minipage}
  \caption{\label{fig4} A schematic network. The directed edges
  are placed according to the rules described the right table.}\vspace{-10pt}
\end{figure}

We consider the schematic network depicted in Fig.~\ref{fig4}(a).
This network is constructed according to the rules in
Fig.~\ref{fig4}(b). Intuitively, according to the outgoing edges in
this network, the nodes can be divided into two groups:
$\{1,2,3,4\}$ and $\{5,6,7,8\}$. Meanwhile, according to the
incoming edges, the nodes of this network belong to another two
groups: $\{1,2,5,6\}$ and $\{3,4,7,8\}$.

We apply Newman's model, Ren's model and our model to this schematic
network. Limited by the assumptions of models, both Newman's model
and Ren's model fail to uncover the intrinsic structural regularity
according the construction rules. For our model, the flexibility of
model assumption enables it to accurately detect such type of
structure. Specifically, by fitting our model to this network, the
obtained $\theta$ reveal the two groups indicated by the outgoing
edges while the $\phi$ reflect the two groups indicated by the
incoming edges.

\subsection{Model selection issue}

In the previous tests, we need to specify the group number before
fitting our model to network. However, the group number is unknown a
prior for many cases. Thus it is helpful to give a criterion to
determinate the appropriate group number for given network. This
task is known as the model selection issue in statistics. We deal
with this problem by using minimum description length principle,
which is also used to handle the model selection issue in Ren's
model.

According to minimum description length principle, the required
length to describe the network data is composed of two parts. The
first part describes the coding length of the network using our
model. This coding length is $-L$ for directed network and $-L/2$
for undirected network. The second part gives the length for coding
model parameters. This part is
$-\sum_{rs}\text{ln}\omega_{rs}-\sum_{ri}(\text{ln}\theta_{ri}+\text{ln}\phi_{ri})$
for directed network and $-\sum_{rs}\text{ln}\omega_{rs}
-\sum_{ri}\text{ln}\theta_{ri}$ for undirected network. In this way,
the optimal $c$ is the one which minimizes the total description
length.

As tests, we consider two real world networks with prior knowledge
of the intrinsic group numbers. These two networks are respectively
the journal citation network constructed in Ref.~\cite{Rosvall07}
and the American football team network described in
Ref.~\cite{Girvan02}. In the journal citation network, each node
corresponds to a journal and all the $40$ journals are from four
different fields: multidisciplinary physics, chemistry, biology and
ecology. Journals from the same field are more likely connected by
citation relation. For the football network, nodes represent the
$115$ teams respectively belonging to $12$ conference and generally
games are more frequent between members of the same conference than
between teams of different conferences.

\begin{figure}
\includegraphics[width=0.48\textwidth]{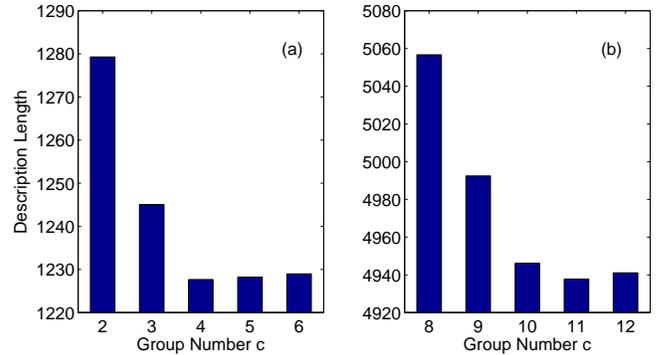}
\caption{Model selection results for (a) the journal citation
network and (b) American football team network.}
\label{fig5}\vspace{-10pt}
\end{figure}

As shown in Fig.~\ref{fig5}, the number of intrinsic groups is
correctly identified for the journal citation network. However, for
the football network, $11$ is the optimal number of groups while the
intrinsic number is $12$. By checking the found node groups, we find
that only $11$ node groups have their identities, i.e., each group
contains at least one node after assigning nodes to their most
preferred groups according to the obtained belonging coefficient
$\alpha$ or $\beta$. This indicates that the appropriate group
number is $11$ for the football network. In fact, many well-known
community detection methods also identify $11$ communities.

\section{Conclusions}
\label{sec4}

In this paper, we have studied the exploration of intrinsic
structural regularities in network using a general stochastic
blockmodel. Without prior knowledge, our model can not only detect
broad types of intrinsic structural regularities, but also can learn
the type of identified structure directly from the network data.
Tests on a number of artificial and real world networks demonstrate
that our model outperforms the state-of-the-art models at shedding
light on the structural features of networks. The flexibility
enables our model be an effective way to reveal the structural
regularities of network and further to help us understand the
relationship between structure and function of network. For
potential application, our model can be used to predict the
emergence or vanishing of edges in network. As future work, we will
generalize our model by releasing the requirement that the block
matrix is a square matrix and investigate the possible applications
of the more flexible model.

\begin{acknowledgments}
This work was funded by the National Natural Science Foundation of
China under grant number $60873245$, $60933005$, $60873217$ and
$60803123$. This work was also partly funded by the National
High-Tech R\&D Program of China (the 863 program) with grant number
2010AA012500. The authors thank Alex Arenas, Mark Newman, and Santo
Fortunato for providing network and other data used in this paper.
\end{acknowledgments}

\end{document}